# CONSeg: Voxelwise Glioma Conformal Segmentation

**Danial Elyassirad[1], Benyamin Gheiji[1], Mahsa Vatanparast[1], Amir Mahmoud Ahmadzadeh, MD[2], Shahriar Faghani, MD[3]***

(1) Student Research Committee, Mashhad University of Medical Sciences, Mashhad, Iran

(2) Department of Radiology, Mashhad University of Medical Sciences, Mashhad, Iran

(3) Radiology Informatics Lab, Department of Radiology, Mayo Clinic, Rochester, Minnesota

(*) Correspondence: Shahriar Faghani, Email: Faghani.Shahriar@mayo.edu

## *Abstract*

**Background and Purpose:** Glioma segmentation is crucial for clinical decisions and treatment planning. Uncertainty quantification methods, including conformal prediction (CP), can enhance segmentation models reliability. This study aims to use CP in glioma segmentation.

**Methods:** We used the UCSF and UPenn glioma datasets, with the UCSF dataset split into training (70%), validation (10%), calibration (10%), and test (10%) sets, and the UPenn dataset divided into external calibration (30%) and external test (70%) sets. A UNet model was trained, and its optimal threshold was set to 0.5 using prediction normalization. To apply CP, the conformal threshold was selected based on the internal/external calibration nonconformity score, and CP was subsequently applied to the internal/external test sets, with coverage reported for all. We defined the uncertainty ratio (UR) and assessed its correlation with the Dice score coefficient (DSC). Additionally, we categorized cases into certain and uncertain groups based on UR and compared their DSC. We also evaluate the correlation between UR and DSC of the BraTS fusion model segmentation (BFMS), and compare DSC in the certain and uncertain subgroups.

**Results:** The base model achieved a DSC of 0.8628 and 0.8257 on the internal and external test sets, respectively. The CP coverage was 0.9982 for the internal test set and 0.9977 for the external test set. Statistical analysis showed a significant negative correlation between UR and DSC for test sets ($p<0.001$). UR was also linked to significantly lower DSCs in the BFMS ($p<0.001$). Additionally, certain cases had significantly higher DSCs than uncertain cases in test sets and the BFMS ($p<0.001$).

**Conclusion:** CP effectively quantifies uncertainty in glioma segmentation. Using CONSeg improves the reliability of segmentation models and enhances human-computer interaction.

## INTRODUCTION

Gliomas are the most prevalent primary brain tumors (1). Accurate segmentation of glioma from MR images is critical for clinical decision-making, including surgical and radiotherapy planning, treatment monitoring, and prognosis assessment (2). Additionally, segmentation masks were used as input for deep learning (DL) models in various tasks, including radiogenomics, grading, and survival prediction (3-5).

DL has advanced glioma segmentation from MRI scans, with the annual BraTS challenge underscoring its growing importance in the field (6,7). Despite these advancements, the robustness of these models remains a concern, and challenges persist regarding their clinical acceptability and reliability across diverse patient populations (8).

One potential approach to address the reliability issue of DL models is uncertainty quantification (UQ) which provides a probabilistic measure of model confidence, enabling the identification of uncertain predictions and improving decision-making in clinical applications (9,10). Several studies have explored uncertainty estimation in image segmentation, aiming to provide more trustworthy predictions (11-13). Among the various techniques, conformal prediction (CP) has emerged as a promising method for UQ (14).

CP quantifies uncertainty in model predictions by generating a set of possible classes for each prediction with a statistical guarantee. This set is constructed to ensure that the label is contained within it with a specified confidence level, thereby providing a rigorous measure of uncertainty in the model's outputs (15). There have been works on the application of CP in image segmentation, highlighting its potential to improve model reliability by providing rigorous uncertainty estimates (16-18). However, to the best of our knowledge, no research has yet applied CP to glioma segmentation.

This study aims to integrate CP with DL-based glioma segmentation to quantify uncertainty in segmentation outputs. By applying CP, we aim to enhance the reliability of the model by identifying uncertain regions in the segmentation, thus improving the model's robustness and clinical applicability. This approach provides a more reliable framework for decision-making in neuro-oncology, especially when dealing with challenging cases.

## METHOD

### Dataset

The UCSF glioma dataset was used for training, validation, calibration, and test set (19). This dataset consists of MRI scans acquired using a 3-T scanner and includes preoperative scans of glioma patients. The inclusion criteria required the presence of the FLAIR sequence and manual segmentation masks for the tumoral area. For model evaluation, we used the UPenn glioma dataset, which followed the same inclusion criteria as the UCSF dataset with the additional criterion that cases also had automated segmentation masks generated using a fusion of top-ranked BraTS Challenge models (20). The detailed acquisition parameters for each dataset have been reported in the original publications (19,20). All images underwent skull stripping and were resampled to 1 mm³ isotropic voxels as part of the original dataset preprocessing.

**Data Preprocessing and splitting**

The UCSF dataset was divided into training (70%), validation (10%), calibration (10%), and test (10%) sets, while the UPenn dataset was split into external calibration (30%) and external test (70%) sets at the patient level. Preprocessing involved background cropping, followed by symmetric zero-padding. Each image was independently normalized, using all nonzero voxels to compute the minimum and maximum voxel values.

**DL Models Development**

We trained a UNet model with FLAIR sequences as input (21). The loss function combined a summation of Dice loss and weighted binary cross-entropy (BCE), where the BCE weight was determined based on the training label distribution.

The best model was selected based on the validation set Dice score coefficient (DSC). To determine the base model optimal threshold (BMOT) for voxel classification, DSCs were evaluated across thresholds from 0 to 1 (in 1% increments), selecting the threshold that maximized the validation DSC. The final model performance was assessed using the DSC across validation, internal, and external test sets, applying the BMOT.

**Conformal Prediction**

To apply CP, we normalized base model voxels' prediction using the following equations (equations 1):

$$Normalized\ prediction\ (If\ voxel > optimum\ threshold) = \frac{Y\ pred - (Optimum\ threshold \pm \varepsilon)}{2 \times (Y \max - ((Optimum\ threshold \pm \varepsilon))} + 0.5$$

$$Normalized\ prediction\ (If\ voxel < optimum\ threshold) = \frac{Y\ pred - (Optimum\ threshold \pm \varepsilon)}{2 \times ((Optimum\ threshold \pm \varepsilon) - Ymin)} + 0.5$$

$$if\ optimum\ threshold\ is\ 1, put - \varepsilon;\ if\ optimum\ threshold\ is\ 0, use + \varepsilon$$

This procedure yielded normalized predictions with an optimal threshold of 0.5.

CP aims to satisfy this equation (equation 2):

$$1 - \alpha \leq P(Y\ true\ \epsilon\ C(xi)) \leq 1 - \alpha + \frac{1}{n+1}$$

Where $\alpha$ is the error (was set to 0.002 in our study), $Y\ true$ is the label of y, $C(xi)$ indicates the prediction set, and $n$ represents the number of cases (in our study, voxels) in the calibration set. Using the calibration set(s), we computed nonconformity scores based on hinge loss (equation 3):

$$Each\ voxel's\ nonconformity\ score = 1 - P(Y\ true)$$

Where the model prediction for the label is shown as $P(Y\ true)$. Subsequently, the nonconformity score threshold (NCST) is determined using the selected α value. To apply CP, for each test voxel, the nonconformity score is computed for both potential classes. A class is included in the prediction set if its corresponding nonconformity score is below the NCST. If neither or both classes are included in the prediction set, the voxel is classified as uncertain. If only one class is included in the prediction set, the voxel is considered certain. Additionally, we report the coverage, defined as the proportion of voxels for which the label is included in the prediction set ($P(Y\ true\ \epsilon\ C(xi))$).

**Conformal Model Validation**

To validate the output of the CP model, the following steps were conducted

1. Uncertainty ratio (UR) analysis for internal test set: The UR for each internal test case was computed, defined as (equation 4):

$$Uncertainity\ ratio = \frac{Number\ of\ uncertain\ voxels}{Number\ of\ certain\ class\ 1\ voxels}$$

Where certain class 1, is the CP model prediction. Next, the correlation coefficient (and corresponding p-value) between the UR of each case in the internal test set and the corresponding model's DSC was assessed.

2. UR analysis for external test set: The UR for external test cases was calculated, and its correlation coefficient (and corresponding p-value) with the DSCs was statistically analyzed.

3. UR clustering for internal model validation: The CP model was applied to the internal calibration set, and the UR was calculated. A K-means clustering algorithm was used to categorize cases based on their UR into two certain and uncertain subgroups. The internal test set was then categorized using the clustering model threshold, and the DSC of these groups were statistically compared.

4. UR clustering for external model validation: To assess generalizability, the same approach used for the internal calibration and test sets was repeated for the external calibration and test sets, including UR calculation, K-means clustering, and statistical comparison of DSC across categorized groups.

5. Comparative analysis with BraTS fusion model segmentation (BFMS) masks: To validate the model's ability to identify uncertain cases, the UR of each external test case was compared with the DSC of automated segmentation masks derived from a fusion of top-ranked BraTS models, and statistical significance was assessed.

6. Threshold-based categorization of external test set, using BFMS masks: Finally, the selected threshold from the external calibration set (using the K-means clustering algorithm) was used to categorize the external test cases, and their DSC were statistically compared against the BFMS masks to further validate the reliability of the CP model.

**Statistical Analysis**

Categorical data were compared using the chi-square test or Fisher's exact test and were reported as frequency (percentage). Normality was assessed using the Kolmogorov-Smirnov test. Quantitative variables were expressed as mean ± standard deviation or median (interquartile range), depending on data distribution. Based on normality, comparisons were conducted using the t-test or Mann-Whitney U test. Correlations between variables were analyzed using the Pearson or Spearman rank correlation test, depending on the data distribution. All statistical analyses were performed using SPSS 20 software, with p < 0.05 considered statistically significant.

## RESULTS

The UCSF dataset comprised 495 cases and the UPenn dataset included 147 cases. The data characteristics summarized in Table 1.

**Table 1:** Data characteristics

|                     | UCSF          | UPenn             | P value |
|---------------------|---------------|-------------------|---------|
| **Number of patients** | 495        | 147               |         |
| **WHO CNS Grade**   |               |                   | <0.001  |
| 2                   | 56(11.31%)    | 0                 |         |
| 3                   | 43(8.69%)     | 0                 |         |
| 4                   | 396(80%)      | (100%)            |         |
| **Sex**             |               |                   | 0.989   |
| Male                | 296(59.8%)    | 88(59.86%)        |         |
| Female              | 199 (40.2%)   | 59 (40.14%)       |         |
| Age                 | 59 (47-68)    | 62.22(53.5-70.14) | <0.001  |

### Segmentation Model Performance

The best model achieved a validation DSC of 0.8745, and a test DCS of 0.8628 on UCSF dataset and DSC of 0.8257 on UPenn dataset. The BMOT was found to be almost 1 (Figure S1).

### Conformal Model Performance

#### UCSF dataset

The CP model achieved a NCST of 0.8874 with coverage of 0.9982 in the UCSF test set. A significant correlation was observed between the UR and the DSC for the UCSF test set (p <0.001, correlation coefficient = -0.827) (Figure S2-3).

The test cases were categorized into uncertain and certain subgroups using a UR threshold of 0.0314 (Figure S4-5). Uncertain cases (20.4%) achieved a DSC of 0.7090, whereas certain cases attained a DSC of 0.9023 (p value < 0.001). The detailed results of this analysis are presented in Table 2.

**Table 2:** DSC for categorized test cases in the UCSF test set

| UR Threshold   | Number of Cases | Mean DSC                               | p-value |
|----------------|-----------------|----------------------------------------|---------|
| UR > 0.0314    | 10 (20.4%)      | 0.7090 (median= 0.77, 0.6747-0.8286)   | < 0.001 |
| UR ≤ 0.0314    | 39 (79.6%)      | 0.9023 (median= 0.9176, 0.8903-0.9357) |         |

**UPenn dataset:**

The CP model reached a NCST of 0.9947 with coverage of 0.9977 in the UPenn test set. A significant correlation was observed between the UR and DSC for the UPenn test set (p <0.001, correlation coefficient = -0.833) (Figure S6).

Next, the cases were categorized based on the UR threshold of 0.12281 (Figure S7-8). A significant difference in DSC between the uncertain (18.4% of cases with DSC of 0.6558) and certain (81.6% of cases, DSC = 0.8642) subgroups was found (p < 0.001). The detailed results of this analysis are presented in Table 3.

**Table 3:** DSC for categorized test cases in the UPenn test set

| UR Threshold | Number of Cases | Mean DSC | p-value |
| --- | --- | --- | --- |
| UR > 0.12281 | 19 (18.45%) | 0.6558 (median= 0.6990, 0.5217-0.7812) | < 0.001 |
| UR ≤ 0.12281 | 84 (81.55%) | 0.8642 (median= 0.8849, 0.8903-0.9357) | |

**Comparative Analysis with BraTS Fusion Model Outputs:**

A significant correlation was found between the UR and DSC of the BFMS (p <0.001, correlation coefficient = -0.562) (Figure S9).

Also, Upenn cases were divided based on the UR threshold of 0.12281. A significant difference in DSC between the uncertain (18.4% cases, DSC = 0.8789) and certain (81.6% cases, DSC = 0.9286) subgroups was observed (p < 0.001). The detailed results of this analysis are presented in Table 4.

**Table 4:** DSC for categorized test cases in the UPenn test set, using BFMS outputs

| UR Threshold | Number of Cases | Mean DSC | p-value |
| --- | --- | --- | --- |
| UR > 0.12281 | 19 (18.45%) | 0.8789 (median= 0.8758, 0.8465-0.9101) | < 0.001 |
| UR ≤ 0.12281 | 84 (81.55%) | 0.9286 (median= 0.9406, 0.9216-0.9517) | |

## *DISCUSSION*

In this study, we demonstrate that conformal segmentation (CONSeg) enhances model reliability by identifying uncertain voxels and cases. We define the UR and find out that it has a significant negative correlation with the DSC in both the internal and external test sets. Additionally, the UR for each case (based on our U-Net model) also exhibits a significant correlation with the DSC of BFMS masks in the external test set. Using the internal and external calibration sets, we determined a threshold for the UR to categorize cases to certain and uncertain cases. Applying this threshold to the internal and external test sets revealed a significant difference in DSC between uncertain and certain cases. Finally, applying the selected threshold to the BFMS masks resulted in a significant difference in the DSC.

In this study, before applying CP, we use a base model prediction normalization (BMPN) process that adjusts the BMOT to 50% and guarantees that uncertain predictions cover both sides of the BMOT (Figure 1). The imbalanced BMOT arises due to factors such as weighted loss and imbalanced datasets.

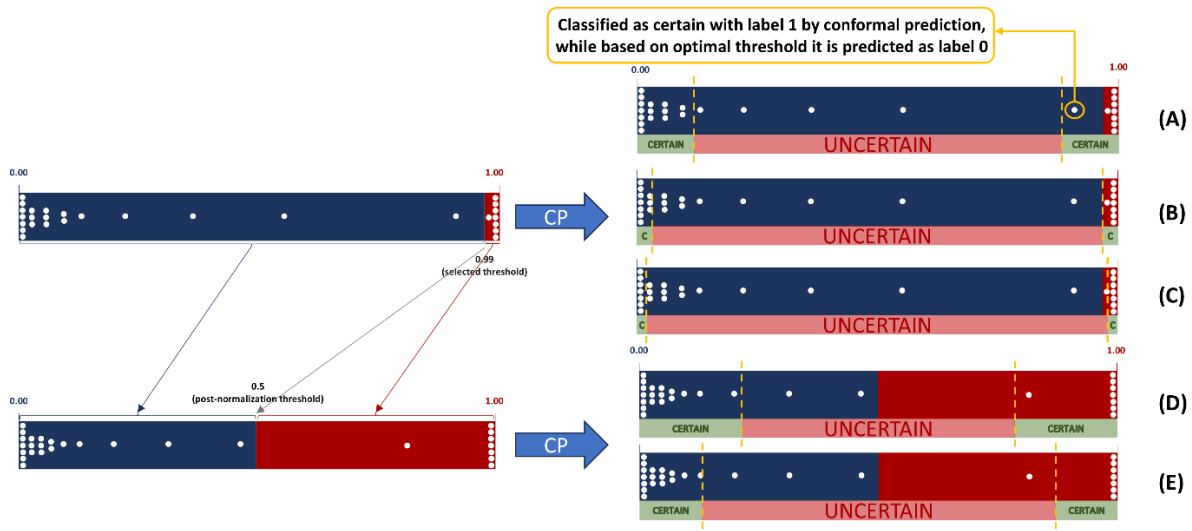

*Figure 1. A-C represent different possible scenarios when applying CP to a model with an imbalanced BMOT, while D and E show the possible scenarios after applying BMPN.*

Figure 1 shows various scenarios of CP with an imbalanced BMOT (e.g., 0.99) and the changes after applying BMPN. Figure 1A shows that when the NCST (or 1- NCST if NCST is below 0.5) is lower than the BMOT, CP classifies some predictions as certain with CP class 1, while those predictions were class 0 based on the BMOT (yellow box in Figure 1A). Figure 2 shows without BMPN, the CP model misclassifies some voxels as CP class 1 when they should be uncertain or certain with CP class 0 (indicated by yellow circles).

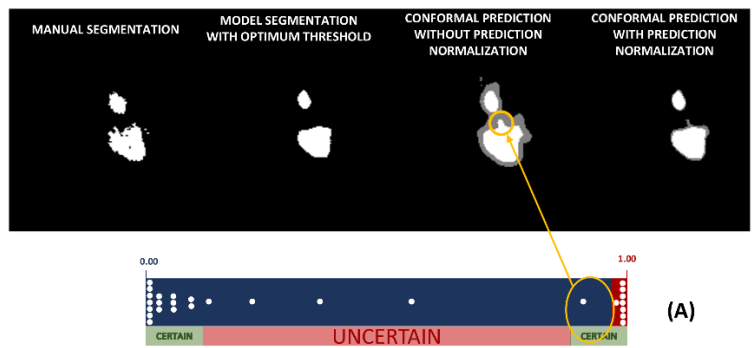

*Figure 2. From left to right; manual segmentation, base model segmentation with BMOT, conformal segmentation without BMPN (gray voxels are uncertain voxels), conformal*

*segmentation with BMPN. This figure shows scenario (A) which happened before BMPN in our CP model.*

Figure 1B, shows what happens when the NCST is close to the BMOT. In this situation, all predictions more than BMOT remain certain with CP class 1, however some predictions below the BMOT are considered uncertain. Also all CP certain classes align with BMOT predictions. Figure 1C represents the ideal scenario, where the NCST is higher than the BMOT, leading to uncertainty in both class 0 and class 1 voxels based on the BMOT. In this case, the CP model's certain classes match with the BMOT classes.

After applying BMPN, Figures 1D and 1E are possible scenarios (both are similar, just in figure 1D no voxel with class 1 based on BMOT categorized as uncertain voxels, like 1B). Overall, BMPN eliminates the misclassification issue observed before normalization (as shown in Figure 1A and 2) and aligns the CP classes more accurately with the BMOT classes.

Few studies have previously applied CP for segmentation. Davenport applied CP for polyp tumor segmentation (17), while Gade et al. used CP for prostate segmentation (18). Based on these studies and our own findings, we have gained a clearer understanding of the advantages that uncertainty quantification in segmentation provides. It serves as a guide for manual correction of model-generated segmentations and helps identify regions more susceptible to misclassification (17,18). Additionally, when the UR is high, it indicates that automated segmentation may not be suitable for the case, necessitating a manual approach. Furthermore, our findings show that uncertain cases, as identified by our model, exhibit significantly lower DSC compared to certain cases, even when utilizing a combination of the best valid segmentation models (BraTS models). This highlights a crucial insight: they are inherently challenging to segment with confidence. Therefore, for such uncertain cases, manual segmentation remains the preferred approach. By incorporating CONSeg, we can enhance human-computer interaction and achieve more reliable segmentation results. CONSeg can be useful in various clinical contexts, including tumor follow-up, assessment of tumor response to therapy, and prognosis. It is also valuable in metastasis detection, as new metastases may not exhibit the typical tumoral features on MRI, making them harder to predict accurately with DL models. CP can help identify these suspicious regions and flag them as uncertain.

Additionally, CONSeg can be used more generally, for tasks like anomaly detection, though this was not the focus of our study. On the other hand, CONSeg can be applied across a wide range of tasks due to the flexibility of CP. It is post-hoc process, model-agnostic, requires minimal computational resources, and does not depend on large datasets, making it a versatile tool for various tasks.

This study is the first to apply CP for segmentation in neuroimaging. Our work introduces several novel contributions, including defining the UR and categorizing segmentation cases into certain and uncertain. Additionally, we implemented prediction normalization, which has not been previously used in CP. However, our study has some limitations. We trained our model just on a single MRI sequence (FLAIR), and all datasets used were publicly available, with model training conducted at a single center. Moreover, k-means clustering may not define the optimal threshold for categorizing UR. Future studies should focus on further evaluating conformal segmentation and incorporating local datasets to enhance the model's generalizability. Additionally, utilizing more MRI sequence(s) may further improve the model's performance.

## *CONCLUSIONS*

CP effectively quantifies uncertainty in glioma segmentation, flagging uncertain voxels and cases. By employing conformal segmentation, we enhance model reliability and human-computer interaction. It also identifies less reliable segmentation cases, recommending them for manual segmentation.

## *CONFLICT OF INTERESTS*

The authors declare no conflicts of interest related to the content of this article.

**SUPPLEMENTAL FILES**

Figure S1. DSC of the segmentation model on the validation set at different thresholds

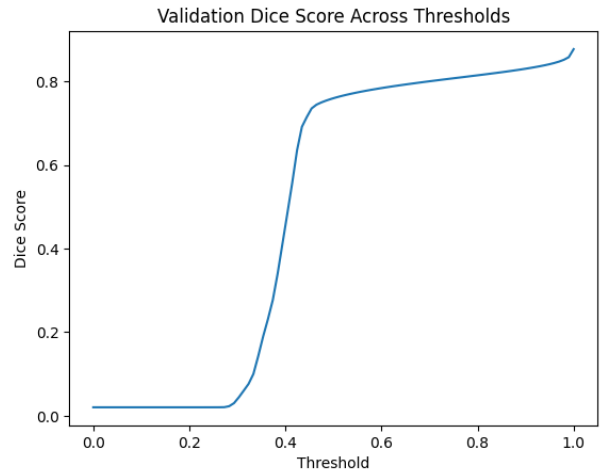

Figure S2. Scatter plot showing the correlation between UR and the DSC on the internal test set (UCSF test set).

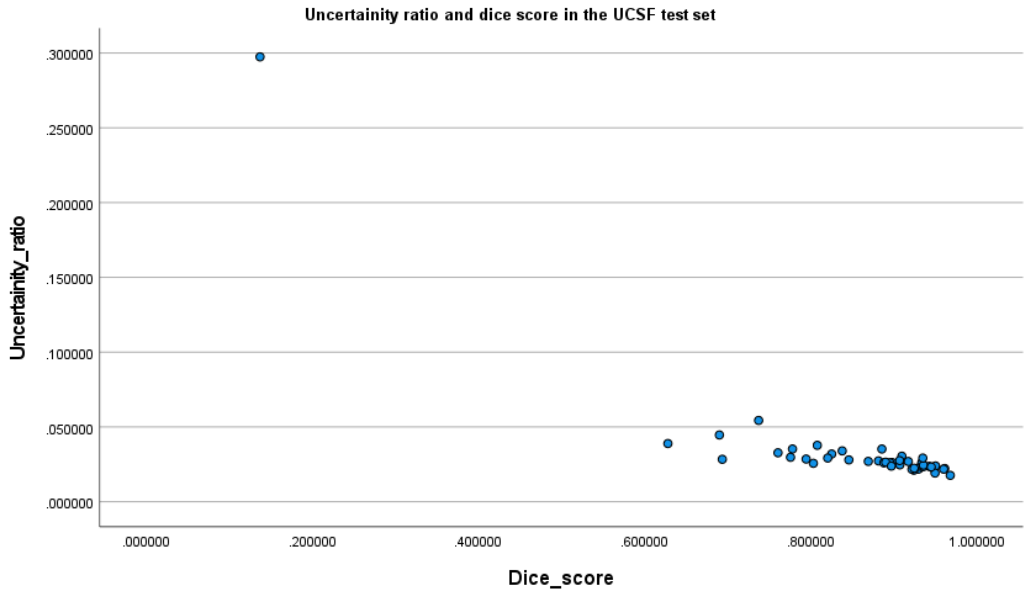

Figure S3. Scatter plot showing the correlation between UR and the DSC on the internal test set (UCSF test set), with outlier(s) removed to better highlight the correlation in cases with lower UR.

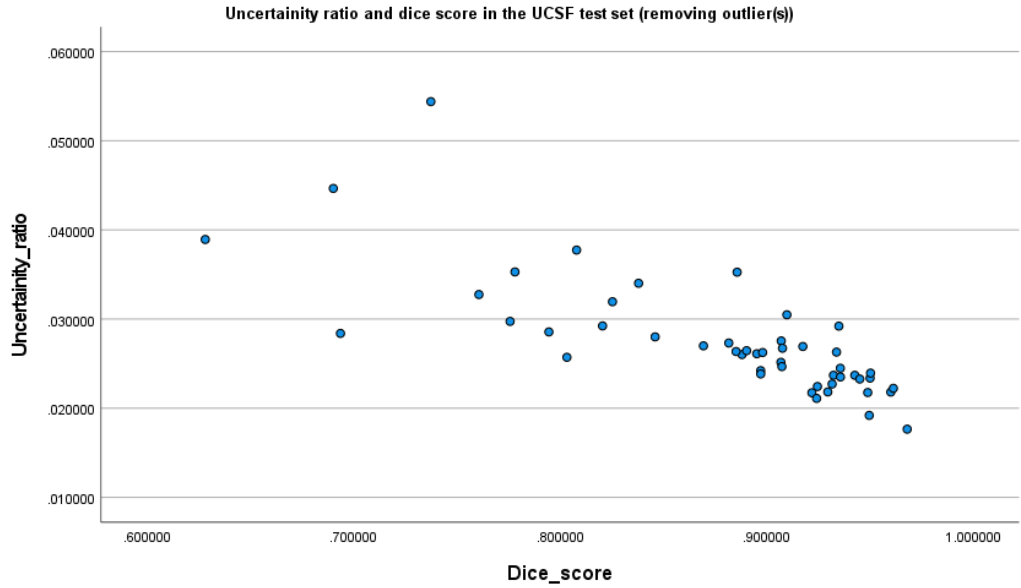

Figure S4. Scatter plot showing the correlation between UR and the DSC on the internal calibration set (UCSF calibration set).

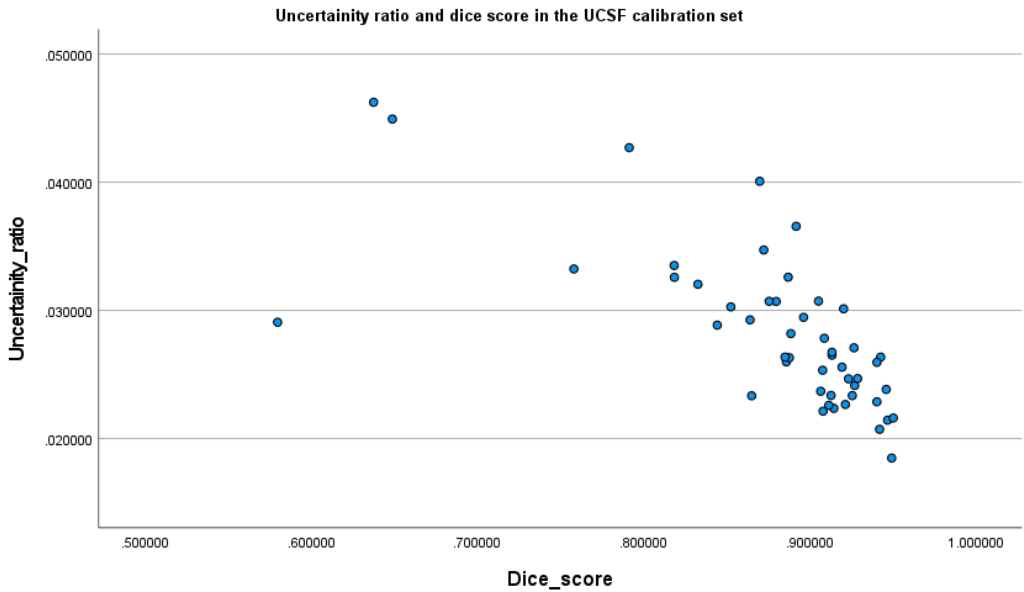

Figure S5. K-means clustering used to determine the threshold on the internal calibration set UR

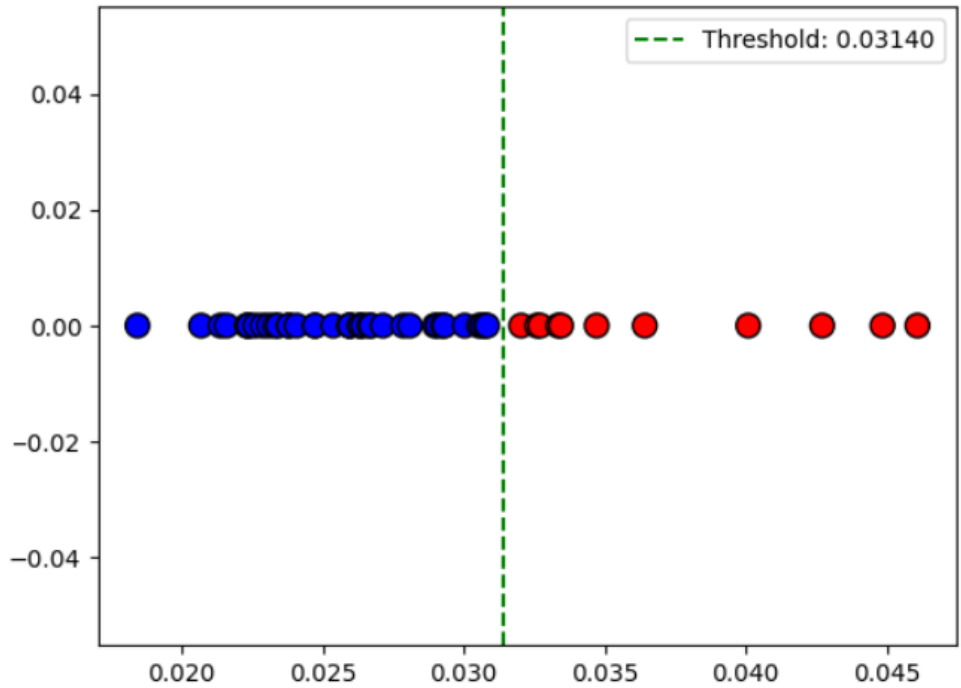

Figure S6. Scatter plot showing the correlation between UR and the DSC on the external test set (UPenn test set).

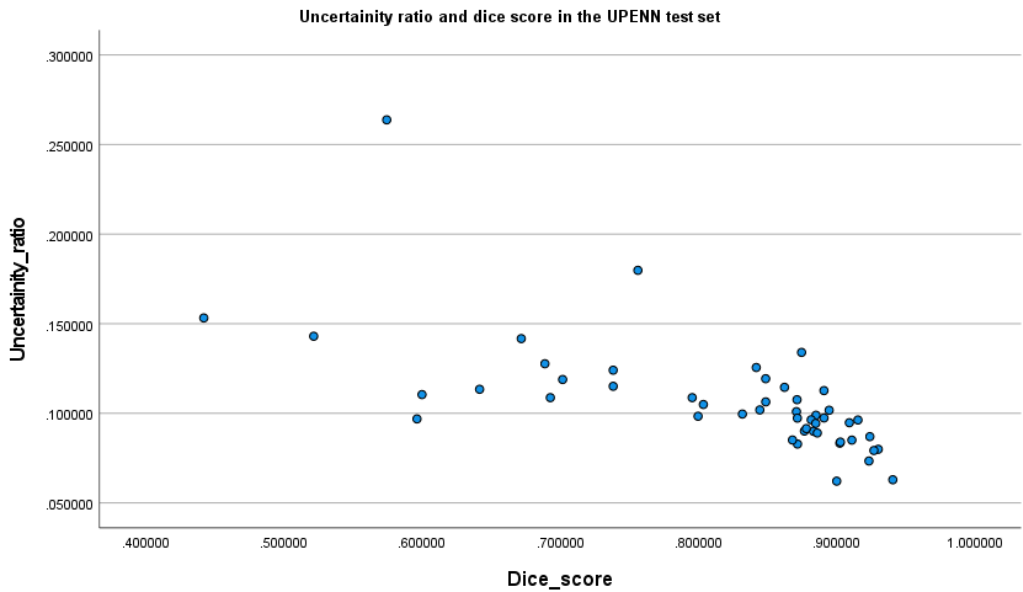

Figure S7. Scatter plot showing the correlation between UR and the DSC on the external calibration set (UPenn calibration set).

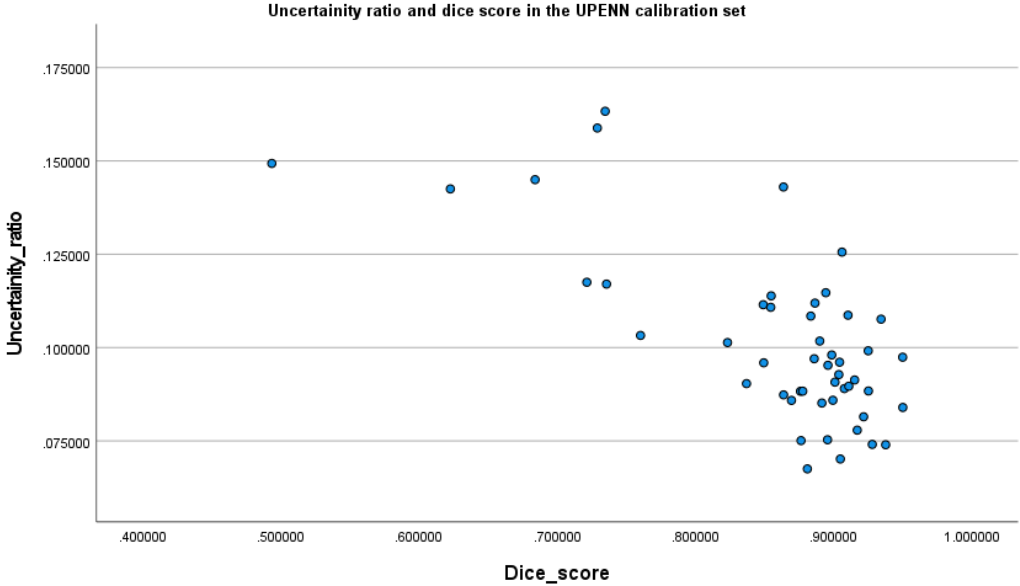

Figure S8. K-means clustering used to determine the threshold on the external calibration set UR

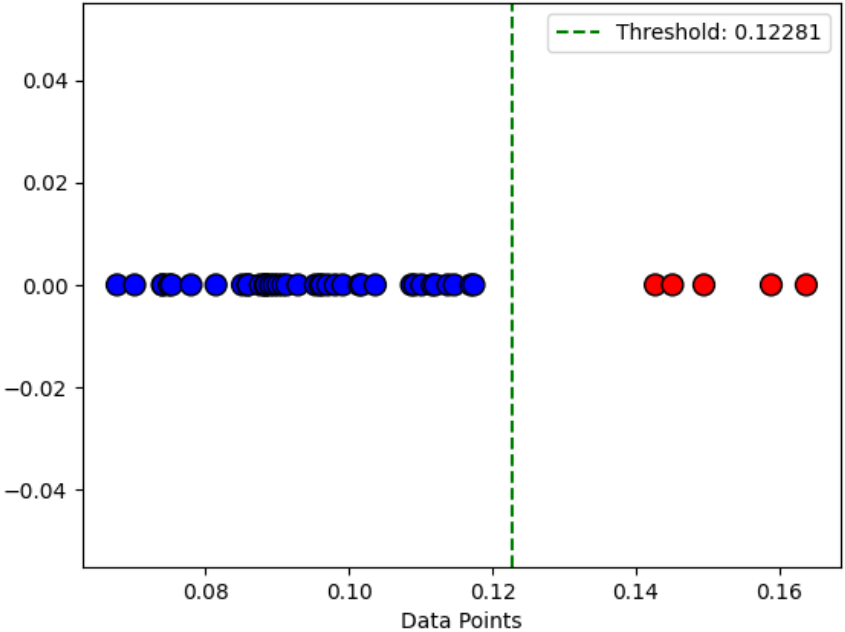

Figure S9. Scatter plot showing the correlation between UR and the BraTS fusion model DSC on the external test set (UPenn test set).